\documentclass[12pt]{article}
\usepackage{amsmath,amssymb,slashed,nicefrac}

\newcommand{\agt}{\rlap{\lower 3.5 pt \hbox{$\mathchar \sim$}} \raise 1pt
 \hbox {$>$}}
\newcommand{\alt}{\rlap{\lower 3.5 pt \hbox{$\mathchar \sim$}} \raise 1pt
 \hbox {$<$}}
\newcommand{\re}{\mathop{\mathrm{Re}}\nolimits}
\newcommand{\im}{\mathop{\mathrm{Im}}\nolimits}
\newcommand{\adj}{\mathop{\mathrm{adj}}\nolimits}

\catcode`@=11
\newcount\@tempcntc
\def\@citex[#1]#2{\if@filesw\immediate\write\@auxout{\string\citation{#2}}\fi
  \@tempcnta\z@\@tempcntb\m@ne\def\@citea{}\@cite{\@for\@citeb:=#2\do
    {\@ifundefined
       {b@\@citeb}{\@citeo\@tempcntb\m@ne\@citea\def\@citea{,}{\bf
?}\@warning
       {Citation `\@citeb' on page \thepage \space undefined}}%
    {\setbox\z@\hbox{\global\@tempcntc0\csname b@\@citeb\endcsname\relax}%
     \ifnum\@tempcntc=\z@ \@citeo\@tempcntb\m@ne
       \@citea\def\@citea{,}\hbox{\csname b@\@citeb\endcsname}%
     \else
      \advance\@tempcntb\@ne
      \ifnum\@tempcntb=\@tempcntc
      \else\advance\@tempcntb\m@ne\@citeo
      \@tempcnta\@tempcntc\@tempcntb\@tempcntc\fi\fi}}\@citeo}{#1}}
\def\@citeo{\ifnum\@tempcnta>\@tempcntb\else\@citea\def\@citea{,}%
  \ifnum\@tempcnta=\@tempcntb\the\@tempcnta\else
   {\advance\@tempcnta\@ne\ifnum\@tempcnta=\@tempcntb \else
\def\@citea{--}\fi
    \advance\@tempcnta\m@ne\the\@tempcnta\@citea\the\@tempcntb}\fi\fi}
\catcode`@=12
\begin{document}

\title{
\vskip-3cm{\baselineskip14pt
\centerline{\normalsize DESY 14-007\hfill ISSN 0418-9833}
\centerline{\normalsize May 2014\hfill}}
\vskip1.5cm
Propagator mixing renormalization for Majorana fermions}

\author{Bernd A. Kniehl
\\
{\normalsize II. Institut f\"ur Theoretische Physik, Universit\"at Hamburg,}
\\
{\normalsize Luruper Chaussee 149, 22761 Hamburg, Germany}
}

\date{}

\maketitle

\begin{abstract}
We consider a mixed system of unstable Majorana fermions in a general
parity-nonconserving theory and renormalize its propagator matrix to all orders
in the pole scheme, in which the squares of the renormalized masses are
identified with the complex pole positions and the wave-function
renormalization matrices are adjusted in compliance with the
Lehmann--Symanzik--Zimmermann reduction formalism.
In contrast to the case of unstable Dirac fermions, the wave-function
renormalization matrices of the in and out states are uniquely fixed, while
they again bifurcate in the sense that they are no longer related by
pseudo-Hermitian conjugation.
We present closed analytic expressions for the renormalization constants in
terms of the scalar, pseudoscalar, vector, and pseudovector parts of the
unrenormalized self-energy matrix, which is computable from the
one-particle-irreducible Feynman diagrams of the flavor transitions, as well
as their expansions through two loops.
In the case of stable Majorana fermions, the well-known one-loop results are
recovered.

\medskip

\end{abstract}

\newpage

\section{Introduction} 

The Standard Model (SM) of elementary particle physics has been enormously
consolidated by the discovery \cite{Aad:2012tfa} at the CERN Large Hadron
Collider of a new weak neutral resonance that, within the present experimental
precision, shares the spin, parity, and charge-conjugation quantum numbers
$J^{PC}=0^{++}$ and the coupling strengths with the SM Higgs boson $H$, in the
complete absence of any signals of new physics beyond the SM.
Moreover, its mass of $(125.6\pm0.3)$~GeV lies well inside the $M_H$ range
predicted within the SM through global analyses of electroweak (EW) precision
data, and it almost perfectly coincides with state-of-the-art determinations,
based on three-loop evolution and two-loop matching, of the $M_H$ lower bound,
$(129.6\pm1.5)$~GeV \cite{Bezrukov:2012sa,Degrassi:2012ry}, from the
requirement that the SM vacuum be stable way up to the scale of the Planck
mass \cite{Krasnikov:1978pu}.
If the pole mass $m_t$ of the top quark, which, in want of a rigorous
determination at the quantum level, is presently identified with a Monte-Carlo
parameter \cite{Beringer:1900zz}, were just lower by an amount of the order of
its total decay width $\Gamma_t=(2.0\pm0.5)$~GeV \cite{Beringer:1900zz},
then the agreement would be perfect, implying that EW symmetry breaking is
likely to be determined by Planck-scale physics \cite{Bezrukov:2012sa}.
In a way, this would solve the longstanding hierarchy problem of the SM.
The Nobel Prize in Physics 2013 was recently awarded jointly to Englert and
Higgs for the theoretical discovery of the Higgs mechanism.

Despite the recent triumph of the SM, we must keep in mind that its neutrino
sector is still holding great longstanding mysteries.
Among the most prominent ones of them is the question whether the neutrinos are
Dirac or Majorana fermions \cite{Gonzalez-Garcia:2013jma}.
In the latter case, physics beyond the SM is indispensable.
On the other hand, numerous beyond-SM scenarios, in particular those in which
the new physics is accessed via a Higgs portal, involve heavy Majorana
neutrinos as ingredients to explain the smallness of the masses of the observed
neutrinos via the seesaw mechanism \cite{Minkowski}.
In the ongoing endeavor to complete the all-order renormalization of the SM and
its most favorable extensions among those not yet excluded experimentally, it
is, therefore, necessary to also accommodate Majorana fermions allowing for
flavor mixing and instability.

The renormalizability of the spontaneously broken quantum gauge theory
underlying the SM was proven in 1971 \cite{'tHooft:1971fh}, and
the Nobel Prize in Physics 1999 was awarded to 't~Hooft and Veltman
{\it for elucidating the quantum structure of EW interactions in physics}.
The on-shell renormalization scheme, which includes the physical particle
masses and Sommerfeld's fine-structure constant among the basic parameters,
provides a natural framework for that.
It was systematically elaborated at one loop for stable particles in
Refs.~\cite{Fleischer:1980ub,Aoki:1982ed,Bohm:1986rj,Hollik:1988ii}, and a
particularly useful variant of it was proposed in Ref.~\cite{Sirlin:1980nh}.
The on-shell renormalization of the SM was established to all orders of
perturbation theory using the algebraic method \cite{Grassi:1995wr}.
However, all particles were assumed to be stable, neutrinos were taken to be
massless, and quark flavor mixing was neglected.
To eliminate these unrealistic assumptions, one needs to develop a pole scheme
of mixing renormalization for unstable particles valid to all orders.
Apart from being conceptually desirable, this is becoming of major
phenomenological importance, both in the SM and beyond, even more so because
mixing and instability of elementary particles prevail and concur in nature.
This requires generalized concepts for flavor-changing propagators and
vertices.
In the SM with massless neutrinos, these are the propagator matrices of the up-
and down-type quarks and their charged-current vertices, which involve the
Cabibbo--Kobayashi--Maskawa (CKM) \cite{Cabibbo:1963yz} quark mixing matrix.
This pattern carries over to the lepton sector if the neutrinos are massive
Dirac fermions, and the analog of the CKM matrix is the
Pontecorvo--Maki--Nakagawa--Sakata \cite{Pontecorvo:1957cp} neutrino mixing
matrix.
Things are more complicated in the presence of Majorana degrees of freedom in
the neutrino sector, which typically give rise to flavor-changing vertices
involving the $Z^0$ and Higgs bosons, too.

As for the renormalization of propagator matrices of mixed systems of
fermions, the situation is as follows.
In Ref.~\cite{Donoghue:1979jq}, an early treatment of finite renormalization
effects both for quarks in hadronic bound states and leptons may be found.
In Ref.~\cite{Gambino:1999ai}, the ultraviolet (UV) renormalization of the
fermion masses was considered, and the pole masses were shown to be gauge
independent to all orders in the SM using Nielsen identities
\cite{Nielsen:1975fs}, both for stable and unstable Dirac fermions.
In Ref.~\cite{Kniehl:2012zb}, the UV renormalization of the Dirac fermion
fields was discussed for the case of stability, and the dressed propagator
matrices were written in closed form, both for the unrenormalized and
renormalized versions.
Furthermore, it was explicitly proven that the wave-function renormalization
(WFR) conditions proposed by Aoki, Hioki, Kawabe, Konuma, and Muta (AHKKM)
\cite{Aoki:1982ed} guarantee the unit-residue properties of the diagonal
elements of the renormalized propagator matrix to all orders, in compliance
with the Lehmann--Symnanzik--Zimmermann (LSZ) reduction formalism
\cite{Lehmann:1954rq}.
In Refs.~\cite{Kniehl:2013uwa,Kniehl:2014}, the discussion of
Ref.~\cite{Kniehl:2012zb} was extended to the case of unstable Dirac fermions,
and closed all-order expressions for their mass conterterms and WFR matrices
were constructed.
The purpose of the present paper is to generalize the approach of
Refs.~\cite{Kniehl:2013uwa,Kniehl:2014} to Majorana fermions.
Specifically, we work out the renormalization of the propagator matrix of
a mixed system of unstable Majorana fermions to all orders.

As for the flavor mixing matrices of Majorana fermions, various renormalization
prescriptions have been proposed at one loop for the case of stability
\cite{Kniehl:1996bd,Diener:2001qt,Almasy:2009kn}.
Specifically, the approach of Ref.~\cite{Almasy:2009kn} is based on
Ref.~\cite{Kniehl:2006bs}.
As pointed out in Ref.~\cite{Diener:2001qt}, necessary conditions for the
renormalized fermion mixing matrices include UV finiteness, gauge independence,
and (pseudo)unitarity.
Furthermore, it is desirable for their counterterms to be on shell, flavor
democratic, finite in the case of fermion mass degeneracy, and expressible in
terms of self-energies only \cite{Kniehl:2009kk}.

This paper is organized as follows.
In Sec.~\ref{sec:two}, we start from the inverse of the unrenormalized
propagator matrix and obtain the dressed propagator matrix by performing the
Dyson resummation \cite{Dyson:1949ha}.
At this point, we define the renormalization conditions for the complex pole
masses in terms of secular equations and solve them to all orders of
perturbation theory.
In Sec.~\ref{sec:three}, we introduce the WFR matrices, explain how they
enter the dressed propagator matrix, and define renormalized self-energies in
such a way that the renormalized propagator matrix emerges from its
unrenormalized counterpart by replacing the unrenormalized self-energies in
the latter by their renormalized counterparts. 
In Sec.~\ref{sec:four}, we generalize the AHKKM WFR conditions
\cite{Aoki:1982ed} to the case of instability, impose them on the inverse of
the renormalized propagator matrix obtained in Sec.~\ref{sec:three}, and
solve them exactly for the WFR matrices, so as to establish them in closed
analytic form valid to all orders of perturbation theory.
In contrast to the Dirac case \cite{Kniehl:2013uwa,Kniehl:2014}, the WFR
matrices are uniquely determined.
A similar observation was made for the case of stability at one loop
\cite{Kniehl:1996bd}.
The generalized AHKKM renormalization conditions also allow us to find an
alternative all-order expression for the pole mass counterterms.
In Sec.~\ref{sec:five}, we demonstrate that WFR bifurcation is an inevitable
consequence of the LSZ condition \cite{Lehmann:1954rq} for unstable Majorana
fermions.
Similar observations were made for unstable Dirac fermions at the one-loop
order \cite{Espriu:2002xv} and to all orders \cite{Kniehl:2013uwa,Kniehl:2014}.
In Sec.~\ref{sec:six}, we expand the all-order expressions for the
renormalization constants derived in Secs.~\ref{sec:two}--\ref{sec:four}
through two loops and cast them into a form ready to use in phenomenological
calculations.
As a by-product, we recover the one-loop results for the case of stability
\cite{Kniehl:1996bd}.
Section~\ref{sec:seven} contains a summary and an outlook.

\section{Unrenormalized dressed propagator matrix}
\label{sec:two}

We consider a system of $N$ unstable Majorana fermions in the context of some
general parity-nonconserving theory with intergeneration mixing.
We denote the bare quantum fields of their flavor eigenstates by
$\psi_i^{\prime0}(x)$, where the subscript $i=1,\ldots,N$ is the flavor index
and the superscript 0 labels bare quantities.
For the sake of a compact notation, we group them into a column vector in
flavor space,
\begin{equation}
\Psi^{\prime0}(x)=\left(\begin{array}{c}\psi_1^{\prime0}(x) \\ \vdots \\
\psi_N^{\prime0}(x)\end{array}\right).
\end{equation}  
The Majorana nature of a fermion to be at the same time its own antiparticle
manifests itself in the condition\footnote{%
Here and in the following, the superscript $T$ implies simultaneous
transposition in the spinor and generation spaces.} 
\begin{equation}
\Psi^{\prime0}(x)=C[\bar{\Psi}^{\prime0}(x)]^T,
\label{eq:majo}
\end{equation}
where $C$ is a unitary matrix in four-dimensional spinor space,
with $C^{-1}=C^\dagger$, which transforms Dirac's $\gamma^\mu$ matrices as
$C\gamma^{\mu T}C^\dagger=-\gamma^\mu$.
From the definition of the $\gamma_5$ matrix,
$\gamma_5=i\gamma^0\gamma^1\gamma^2\gamma^3$, it follows that
$C\gamma_5^TC^\dagger=\gamma_5$.
Because $\bar{\Psi}^{\prime0}(x)=[{\Psi}^{\prime0}(x)]^\dagger\gamma^0$ may
also be evaluated from Eq.~(\ref{eq:majo}) as
$\bar{\Psi}^{\prime0}(x)=[\Psi^{\prime0}(x)]^TC^*$, $C$ must satisfy the
additional condition $C^\dagger=-C^*$, by being antisymmetric, $C^T=-C$.
The kinetic term of the bare Lagrangian is
\begin{equation}
\mathcal{L}^0(x)=\frac{1}{2}[\Psi^{\prime0}(x)]^TC^*(i\slashed{\partial}
-\mathcal{M}^{\prime0})\Psi^{\prime0}(x),
\label{eq:barel}
\end{equation}
where $\mathcal{M}^{\prime0}$ is the bare mass matrix and the overall factor
$\nicefrac{1}{2}$ corrects for the seeming paradox that each Majorana fermion
contributes to the total energy twice, as a particle and an antiparticle.
For $\mathcal{L}^0(x)$ to be Hermitian,
$[\mathcal{L}^0(x)]^*=\mathcal{L}^0(x)$, $\mathcal{M}^{\prime0}$ must satisfy
the pseudo-Hermiticity relation
$\gamma^0\mathcal{M}^{\prime0\dagger}\gamma^0=\mathcal{M}^{\prime0}$, which
constrains it to the form
$\mathcal{M}^{\prime0}=M^{\prime0}a_++M^{\prime0\dagger}a_-$, where
$M^{\prime0}$ is an arbitrary complex $N\times N$ matrix and
$a_\pm=(I_4\pm\gamma_5)/2$ are the chiral projection operators.
Here and in the following, $I_n$ denotes the $n\times n$ unit matrix.
Exploiting the identity $[\mathcal{L}^0(x)]^T=\mathcal{L}^0(x)$ in connection
with the Grassmannian property of fermionic quantum fields, we obtain the
additional relation $C\mathcal{M}^{\prime0T}C^\dagger=\mathcal{M}^{\prime0}$,
which implies that $M^{\prime0}$ is symmetric, $M^{\prime0T}=M^{\prime0}$.
By Autonne--Takagi matrix factorization,\footnote{%
An explicit proof of this theorem may be found, e.g., in Appendix~B of
Ref.~\cite{Almasy:2009kn}.} 
the complex, symmetric $N\times N$ matrix $M^{\prime0}$ may be transformed into
a real, diagonal matrix $M^0$ with nonnegative entries,\footnote{%
In this paper, summation over repeated indices is not implied in the absence
of summation symbols.}
\begin{equation}
M_{ij}^0=m_i^0\delta_{ij},
\end{equation}
via a single unitary matrix $U_-$, as $M^0=U_-M^{\prime0}U_-^T$.
The bare field multiplet of the mass eigenstates $\psi_i^0(x)$, with bare
masses $m_i^0$, is then given by $\Psi^0(x)=U\Psi^{\prime0}(x)$, where
$U=U_-^*a_++U_-a_-$.
Upon this field transformation, Eq.~(\ref{eq:barel}) indeed assumes the
standard form
\begin{eqnarray}
\mathcal{L}^0(x)&=&\frac{1}{2}[\Psi^0(x)]^TC^*(i\slashed{\partial}-M^0)
\Psi^0(x)
\nonumber\\
&=&\frac{1}{2}\{i[\Psi_-^0(x)]^TC^*\slashed{\partial}\Psi_+^0(x)
+i[\Psi_+^0(x)]^TC^*\slashed{\partial}\Psi_-^0(x)
\nonumber\\
&&{}-[\Psi_+^0(x)]^TC^*M^0\Psi_+^0(x)
-[\Psi_-^0(x)]^TC^*M^0\Psi_-^0(x)\},
\end{eqnarray}
where $\Psi_\pm^0(x)=a_\pm\Psi^0(x)$ are the right- and left-handed field
components, respectively.
Owing to the identity $\gamma^0CU^*C^\dagger\gamma^0=U$, the Majorana property
of the weak eigenstates in Eq.~(\ref{eq:majo}) carries over to the mass
eigenstates,
\begin{equation}
\Psi^0(x)=C[\bar{\Psi}^0(x)]^T.
\label{eq:majom}
\end{equation}
However, it does not hold separately for $\Psi_\pm^0(x)$.
Instead, we have $\Psi_\pm^0(x)=C[\bar{\Psi}_\mp^0(x)]^T$ \cite{Kniehl:1996bd}.

In momentum space, the unrenormalized propagator matrix is defined as
\begin{equation}
iP(\slashed{p})=\int d^4x\,e^{ip\cdot x}
\langle0|T\{\Psi^0(x)\otimes[\Psi^0(0)]^TC^*\}|0\rangle,
\label{eq:prop}
\end{equation}
where $T$ is the time-ordered product and $\otimes$ is to indicate a tensorial
product both in the spinor and generation spaces.
Its inverse is built up by the one-particle-irreducible Feynman diagrams
contributing to the transitions $j\to i$ and has the form
\begin{equation}
[P(\slashed{p})]^{-1}=\slashed{p}-M^0-\Sigma(\slashed{p}),
\label{eq:bareinv}
\end{equation}
where $\Sigma(\slashed{p})$ is the unrenormalized self-energy matrix.
By Lorentz covariance, the latter exhibits the structure
\begin{equation}
\Sigma(\slashed{p})=[A_+(p^2)+\slashed{p}B_+(p^2)]a_+
+(+\leftrightarrow-),
\label{eq:sigma}
\end{equation}
where the entries in the matrices $A_\pm(p^2)$ and $B_\pm(p^2)$ are
Lorentz-invariant functions of $p^2$.
The latter may be calculated from the bare Lagrangian order by order in
perturbation theory.
However, we refrain from explicitly performing
perturbative expansions in the
following, rendering our results valid to all orders.
Defining
\begin{equation}
S_\pm(p^2)=I_N-B_\pm(p^2),\qquad T_\pm(p^2)=M^0+A_\pm(p^2),
\label{eq:stbare}
\end{equation}
Eq.~(\ref{eq:bareinv}) may be cast into a compact form,
\begin{equation}
[P(\slashed{p})]^{-1}=[\slashed{p}S_+(p^2)-T_+(p^2)]a_+
+(+\leftrightarrow-).
\label{eq:bareinv1}
\end{equation}
The one-particle-reducible Feynman diagrams may be collected systematically by
performing the Dyson resummation \cite{Dyson:1949ha}, which is equivalent to
inverting Eq.~(\ref{eq:bareinv1}) and yields \cite{Kniehl:2012zb}
\begin{eqnarray}
P(\slashed{p})&=&[\slashed{p}+D_-(p^2)]S_-^{-1}(p^2)[p^2-E_-(p^2)]^{-1}a_+
+(+\leftrightarrow-)
\nonumber\\
&=&a_+[p^2-F_+(p^2)]^{-1}S_+^{-1}(p^2)[\slashed{p}+C_+(p^2)]
+(+\leftrightarrow-),
\label{eq:bare}
\end{eqnarray}
with the short-hand notations
\begin{eqnarray}
C_\pm(p^2)&=&T_\mp(p^2)S_\mp^{-1}(p^2),\qquad
D_\pm(p^2)=S_\mp^{-1}(p^2)T_\pm(p^2),
\nonumber\\
E_\pm(p^2)&=&C_\pm(p^2)C_\mp(p^2),\qquad
F_\pm(p^2)=D_\mp(p^2)D_\pm(p^2).
\label{eq:cdbare}
\end{eqnarray}
In fact, Eqs.~(\ref{eq:bareinv1}) and (\ref{eq:bare}) are easily seen to
satisfy
$P(\slashed{p})[P(\slashed{p})]^{-1}=[P(\slashed{p})]^{-1}P(\slashed{p})
=I_N\otimes I_4$.
From the first equality in Eq.~(\ref{eq:stbare}) it follows that
$\det[S_\pm(p^2)]=1+\mathcal{O}(\alpha)\ne0$ with $\alpha$ being a generic
coupling constant, so that we may evaluate $S_\pm^{-1}(p^2)$ as
$S_\pm^{-1}(p^2)=\adj[S_\pm(p^2)]/\det[S_\pm(p^2)]$.\footnote{\label{foot}%
The adjugate (classical adjoint) $\adj A=C^T$ of a quadratic ($n\times n$)
matrix $A$ is the transpose of the matrix $C$ the elements $C_{ij}$ of which
are the cofactors of the elements $A_{ij}$ of $A$.
The cofactor $C_{ij}$ of the element $A_{ij}$ of $A$ is $(-1)^{i+j}$ times the
determinant of the $(n-1)\times(n-1)$ matrix obtained by deleting the $i$th
row and the $j$th column of $A$.
The theorem $A\adj A=(\adj A)A=(\det A)I_n$ may be understood by observing
that, according to Laplace's expansion formula, $\sum_{k=1}^nA_{ik}C_{jk}$ is
the determinant of the matrix obtained from $A$ by replacing the $j$th row by
the $i$th row and $\sum_{k=1}^nC_{ki}A_{kj}$ is the determinant of the matrix
obtained from $A$ by replacing the $i$th column by the $j$th column.
If $i=j$, then, in both cases, the result is just $\det A$.
If $i\ne j$, then it is zero because these determinants have two identical
rows and columns, respectively.
If $\det A\ne0$, then $A^{-1}=(\det A)^{-1}\adj A$.}
Alternatively, we may compute $S_\pm^{-1}(p^2)$ perturbatively as a geometric
series, $S_\pm^{-1}(p^2)=I_N+\sum_{n=1}^\infty B_\pm^n(p^2)$.

Since the four matrices $[p^2-E_\pm(p^2)]$ and $[p^2-F_\pm(p^2)]$, the inverses
of which appear in the individual propagator parts in Eq.~(\ref{eq:bare}), are
related by similarity transformations,
\begin{eqnarray}
p^2-E_+(p^2)&=&C_-(p^2)[p^2-E_-(p^2)]C_-^{-1}(p^2)
\nonumber\\
&=&S_+(p^2)[p^2-F_+(p^2)]S_+^{-1}(p^2)
\nonumber\\
&=&T_-(p^2)[p^2-F_-(p^2)]T_-^{-1}(p^2),
\end{eqnarray}
we have
\begin{equation}
\det[p^2-E_+(p^2)]=\det[p^2-E_-(p^2)]=\det[p^2-F_+(p^2)]=\det[p^2-F_-(p^2)],
\label{eq:det}
\end{equation}
which, by virtue of footnote~\ref{foot}, guarantees that the individual
propagator parts all have poles at the same (complex) positions $p^2=M_i^2$
defined as the zeros of Eq.~(\ref{eq:det}) by any of the secular equations
\cite{Donoghue:1979jq,Gambino:1999ai,Kniehl:2012zb}
\begin{equation}
\det[M_i^2-E_\pm(M_i^2)]=\det[M_i^2-F_\pm(M_i^2)]=0.
\label{eq:zero}
\end{equation}
Here, $M_i$ is the complex pole mass of Majorana fermion $i$.
It is related to the real pole mass $m_i$ and total decay width $\Gamma_i$ as
\cite{Smith:1996xz,Kniehl:2008cj}
\begin{equation}
M_i=m_i-i\frac{\Gamma_i}{2}.
\label{eq:real}
\end{equation}

In the Appendix of Ref.~\cite{Kniehl:2012zb}, Eq.~(\ref{eq:zero}) is solved
perturbatively through two loops for the case of stable Dirac fermions.
Here, we derive closed all-order expressions for $M_i$ in terms of the
Lorentz-invariant functions $[A_\pm(p^2)]_{ij}$ and $[B_\pm(p^2)]_{ij}$.
Owing to footnote~\ref{foot}, we have the identities
\begin{eqnarray}
\{[p^2-F_\pm(p^2)]\adj[p^2-F_\pm(p^2)]\}_{ii}&=&\det[p^2-F_\pm(p^2)],
\nonumber\\
\{\adj[p^2-E_\pm(p^2)][p^2-E_\pm(p^2)]\}_{ii}&=&\det[p^2-E_\pm(p^2)].
\label{eq:id}
\end{eqnarray}
At this point, we introduce the two matrices,
\begin{equation}
M_{ij}^\pm=\{\adj[M_j^2-F_\pm(M_j^2)]\}_{ij},\qquad
\bar{M}_{ij}^\pm=\{\adj[M_i^2-E_\pm(M_i^2)]\}_{ij},
\label{eq:m}
\end{equation}
which we shall need again later, and observe that
$M_{ii}^\pm,\bar{M}_{ii}^\pm=\prod_{j\ne i}(M_i^2-M_j^2)
+\mathcal{O}(\alpha)\ne0$.
For the solution $p^2=M_i^2$ of Eq.~(\ref{eq:zero}), we then obtain from
Eq.~(\ref{eq:id}) the exact expression
\begin{equation}
M_i^2=\frac{[F_\pm(M_i^2)M^\pm]_{ii}}{M_{ii}^\pm}
=\frac{[\bar{M}^\pm E_\pm(M_i^2)]_{ii}}{\bar{M}_{ii}^\pm}.
\label{eq:mfe}
\end{equation}
We shall see later that the WFR procedure generates yet another closed
all-order expression for $M_i$, namely, the one specified in
Eq.~(\ref{eq:mass}), in which both matrices of Eq.~(\ref{eq:m}) enter in a
symmetric way.

Equations~(\ref{eq:bareinv})--(\ref{eq:mfe}) also apply to unstable Dirac
fermions as they stand and were partly presented in
Refs.~\cite{Kniehl:2013uwa,Kniehl:2014}.
However, there are additional constraints for Majorana fermions.
In fact, making use of the Grassmannian nature of fermionic quantum fields in
Eq.~(\ref{eq:prop}), we find
\begin{equation}
C[P(-\slashed{p})]^TC^\dagger=P(\slashed{p}).
\label{eq:propt}
\end{equation}
Applying Eq.~(\ref{eq:propt}) to Eq.~(\ref{eq:bareinv}), we obtain
$C[\Sigma(-\slashed{p})]^TC^\dagger=\Sigma(\slashed{p})$, which implies via
Eqs.~(\ref{eq:sigma}), (\ref{eq:stbare}), (\ref{eq:cdbare}), and (\ref{eq:m})
that
\begin{eqnarray}
A_\pm^T(p^2)&=&A_\pm(p^2),\qquad
B_\pm^T(p^2)=B_\mp(p^2),\qquad
S_\pm^T(p^2)=S_\mp(p^2),\qquad
T_\pm^T(p^2)=T_\pm(p^2),
\nonumber\\
C_\pm^T(p^2)&=&D_\mp(p^2),\qquad
E_\pm^T(p^2)=F_\mp(p^2),\qquad
(M^\pm)^T=\bar{M}^\mp,
\label{eq:transpose}
\end{eqnarray}
where we have used $\adj(A^T)=(\adj A)^T$, which follows from
footnote~\ref{foot}, in the last equality.
The first equality in Eq.~(\ref{eq:transpose}) agrees with the second equality
in Eq.~(4.7) of Ref.~\cite{Kniehl:1996bd}.
It is interesting to observe that the last equality in Eq.~(\ref{eq:mfe}) may
be derived from the last two equalities in Eq.~(\ref{eq:transpose}) alone,
without recourse to Eq.~(\ref{eq:id}), by noticing that diagonal matrix
elements are invariant under transposition.

\section{Renormalized dressed propagator matrix}
\label{sec:three}

In the following, we adopt the pole renormalization scheme, in which the
complex pole masses $M_i$ serve as the renormalized masses; i.e., the mass
counterterms $\delta M_i$ are fixed by the relations
\begin{equation}
m_i^0=M_i+\delta M_i.
\label{eq:mren}
\end{equation}
The field renormalization is implemented by writing
\begin{equation}
\Psi^0(x)=Z^{\nicefrac{1}{2}}\Psi(x),
\label{eq:wfr}
\end{equation}
where $\Psi(x)$ is the renormalized field multiplet and
\begin{equation}
Z^{\nicefrac{1}{2}}=Z_+^{\nicefrac{1}{2}}a_++Z_-^{\nicefrac{1}{2}}a_-,
\label{eq:wfrc}
\end{equation}
with $Z_\pm^{\nicefrac{1}{2}}$ being the WFR matrices.
From Eq.~(\ref{eq:wfr}), it follows that
\begin{equation}
[\Psi^0(x)]^TC^*=[\Psi(x)]^TC^*\bar{Z}^{\nicefrac{1}{2}},
\label{eq:wfrt}
\end{equation}
where
\begin{equation}
\bar{Z}^{\nicefrac{1}{2}}=CZ^{T\nicefrac{1}{2}}C^\dagger
=a_+Z_+^{T\nicefrac{1}{2}}+a_-Z_-^{T\nicefrac{1}{2}}.
\label{eq:wfrct}
\end{equation}
Solving Eqs.~(\ref{eq:wfr}) and (\ref{eq:wfrt}) for the renormalized field
multiplets, we have
\begin{equation}
\Psi(x)=Z^{-\nicefrac{1}{2}}\Psi^0(x),\qquad
[\Psi(x)]^TC^*=[\Psi^0(x)]^TC^*\bar{Z}^{-\nicefrac{1}{2}},
\label{eq:wfri}
\end{equation}
where
\begin{equation}
Z^{-\nicefrac{1}{2}}=Z_+^{-\nicefrac{1}{2}}a_++Z_-^{-\nicefrac{1}{2}}a_-,\qquad
\bar{Z}^{-\nicefrac{1}{2}}=a_+Z_+^{T-\nicefrac{1}{2}}
+a_-Z_-^{T-\nicefrac{1}{2}}
\end{equation}
are the inverses of the matrices in Eqs.~(\ref{eq:wfrc}) and (\ref{eq:wfrct}),
respectively.
Using Eq.~(\ref{eq:wfri}), we may express the renormalized propagator matrix,
\begin{equation}
i\hat{P}(\slashed{p})=\int d^4x\,e^{ip\cdot x}
\langle0|T\{\Psi(x)\otimes[\Psi(0)]^TC^*\}|0\rangle,
\end{equation}
in terms of the unrenormalized one in Eq.~(\ref{eq:prop}) as
\begin{equation}
\hat{P}(\slashed{p})
=Z^{-\nicefrac{1}{2}}P(\slashed{p})\bar{Z}^{-\nicefrac{1}{2}}.
\label{eq:rel}
\end{equation}
Substituting Eq.~(\ref{eq:bare}) in Eq.~(\ref{eq:rel}), we thus obtain 
\begin{eqnarray}
\hat{P}(\slashed{p})&=&[Z_-^{-\nicefrac{1}{2}}\slashed{p}
+Z_+^{-\nicefrac{1}{2}}D_-(p^2)]S_-^{-1}(p^2)[p^2-E_-(p^2)]^{-1}
Z_+^{T-\nicefrac{1}{2}}a_+
+(+\leftrightarrow-)
\nonumber\\
&=&a_+Z_+^{-\nicefrac{1}{2}}[p^2-F_+(p^2)]^{-1}S_+^{-1}(p^2)
[\slashed{p}Z_-^{T-\nicefrac{1}{2}}
+C_+(p^2)Z_+^{T-\nicefrac{1}{2}}]
+(+\leftrightarrow-).
\label{eq:renz}
\end{eqnarray}
We may absorb the WFR matrices in Eq.~(\ref{eq:renz}) by defining renormalized
counterparts of $S_\pm(p^2)$ and $T_\pm(p^2)$ in Eq.~(\ref{eq:stbare}) as
\begin{equation}
\hat{S}_\pm(p^2)=Z_\mp^{T\nicefrac{1}{2}}S_\pm(p^2)Z_\pm^{\nicefrac{1}{2}},
\qquad
\hat{T}_\pm(p^2)=Z_\pm^{T\nicefrac{1}{2}}T_\pm(p^2)Z_\pm^{\nicefrac{1}{2}}.
\label{eq:st}
\end{equation}
In analogy to Eq.~(\ref{eq:cdbare}), we are thus led to define
\begin{eqnarray}
\hat{C}_\pm(p^2)&=&\hat{T}_\mp(p^2)\hat{S}_\mp^{-1}(p^2)
=Z_\mp^{T\nicefrac{1}{2}}C_\pm(p^2)Z_\pm^{T-\nicefrac{1}{2}},
\nonumber\\
\hat{D}_\pm(p^2)&=&\hat{S}_\mp^{-1}(p^2)\hat{T}_\pm(p^2)
=Z_\mp^{-\nicefrac{1}{2}}D_\pm(p^2)Z_\pm^{\nicefrac{1}{2}},
\nonumber\\
\hat{E}_\pm(p^2)&=&\hat{C}_\pm(p^2)\hat{C}_\mp(p^2)
=Z_\mp^{T\nicefrac{1}{2}}E_\pm(p^2)Z_\mp^{T-\nicefrac{1}{2}},
\nonumber\\
\hat{F}_\pm(p^2)&=&\hat{D}_\mp(p^2)\hat{D}_\pm(p^2)
=Z_\pm^{-\nicefrac{1}{2}}F_\pm(p^2)Z_\pm^{\nicefrac{1}{2}}.
\label{eq:cd}
\end{eqnarray}
Thus, Eq.~(\ref{eq:renz}) becomes
\begin{eqnarray}
\hat{P}(\slashed{p})&=&[\slashed{p}+\hat{D}_-(p^2)]\hat{S}_-^{-1}(p^2)
[p^2-\hat{E}_-(p^2)]^{-1}a_+
+(+\leftrightarrow-)
\nonumber\\
&=&a_+[p^2-\hat{F}_+(p^2)]^{-1}\hat{S}_+^{-1}(p^2)[\slashed{p}+\hat{C}_+(p^2)]
+(+\leftrightarrow-).
\label{eq:ren}
\end{eqnarray}
By observing from the last two lines of Eq.~(\ref{eq:cd}) that
\begin{equation}
\det[p^2-\hat{E}_\pm(p^2)]=\det[p^2-E_\pm(p^2)],\qquad
\det[p^2-\hat{F}_\pm(p^2)]=\det[p^2-F_\pm(p^2)],
\label{eq:rendet}
\end{equation}
we understand that the pole positions $M_i^2$ are not affected by the WFR, as
it should be \cite{Gambino:1999ai}.
{\it Mutatis mutandis}, the inverse of the renormalized propagator matrix reads
\begin{eqnarray}
[\hat{P}(\slashed{p})]^{-1}&=&
[\slashed{p}\hat{S}_+(p^2)-\hat{T}_+(p^2)]a_+
+(+\leftrightarrow-)
\nonumber\\
&=&[Z_-^{T\nicefrac{1}{2}}\slashed{p}S_+(p^2)
-Z_+^{T\nicefrac{1}{2}}T_+(p^2)]Z_+^{\nicefrac{1}{2}}a_+
+(+\leftrightarrow-).
\label{eq:reninv}
\end{eqnarray}

The counterparts of Eqs.~(\ref{eq:wfrt}) and (\ref{eq:wfrct}) for unstable
Dirac fermions read \cite{Kniehl:2013uwa,Kniehl:2014}
\begin{equation}
\bar{\Psi}^0(x)=\bar{\Psi}(x)\bar{Z}^{\nicefrac{1}{2}}
\end{equation}
and
\begin{equation}
\bar{Z}^{\nicefrac{1}{2}}=a_-\bar{Z}_+^{\nicefrac{1}{2}}+a_+\bar{Z}_-^{\nicefrac{1}{2}},
\label{eq:wfrcd}
\end{equation}
while the relationship in the first equality of Eq.~(\ref{eq:wfrct}) does not
hold then.
Nevertheless,
Eqs.~(\ref{eq:renz})--(\ref{eq:ren}) and (\ref{eq:reninv}) may be recovered
from Ref.~\cite{Kniehl:2014} via the substitution
\begin{equation}
\bar{Z}_\pm^{\nicefrac{1}{2}}=Z_\mp^{T\nicefrac{1}{2}},
\label{eq:sub}
\end{equation}
which may be gleaned by comparing Eqs.~(\ref{eq:wfrct}) and (\ref{eq:wfrcd}).

\section{Generalized WFR conditions}
\label{sec:four}

Similarly to the case of unstable Dirac fermions
\cite{Kniehl:2013uwa,Kniehl:2014}, we determine the WFR matrices
$Z_\pm^{\nicefrac{1}{2}}$ by requiring that, when any Majorana fermion $n$
approaches its mass shell, $\slashed{p}\to M_n$, the respective diagonal
element $[\hat{P}(\slashed{p})]_{nn}$ of the renormalized propagator resonates
with unit residue, while all the other elements stay finite, i.e.,
\begin{equation}
[\hat{P}(\slashed{p})]_{ij}=
\frac{\delta_{in}\delta_{nj}}{\slashed{p}-M_n}+\mathcal{O}(1),
\label{eq:unit}
\end{equation}
in accordance with the LSZ reduction formalism \cite{Lehmann:1954rq}.
The behavior of $\hat{P}(\slashed{p})$ in Eq.~(\ref{eq:unit}) necessitates that
$[\hat{P}(\slashed{p})]^{-1}$ behaves as
\begin{equation}
\{[\hat{P}(\slashed{p})]^{-1}\}_{ij}=\left\{
\begin{array}{l@{\quad\mathrm{if}\quad}l}
(\slashed{p}-M_n)[I_4+\mathcal{O}(\slashed{p}-M_n)] & i=n=j, \\
{[M_{in}+\mathcal{O}(\slashed{p}-M_n)]}(\slashed{p}-M_n) & i\ne n=j, \\
(\slashed{p}-M_n)[M_{nj}+\mathcal{O}(\slashed{p}-M_n)] & i=n\ne j, \\
M_{ij}+\mathcal{O}(\slashed{p}-M_n) & i\ne n\ne j,
\end{array}
\right.
\label{eq:aoki0}
\end{equation}
where $M_{ij}$ are constant matrices in spinor space, which, in general, do not
commute with $\slashed{p}$.
In fact, they are linear combinations of the Dirac matrices $I_4$ and
$\gamma_5$ with constant coefficients.
The specific structure of Eq.~(\ref{eq:aoki0}) may be easily understood by
multiplying Eqs.~(\ref{eq:unit}) and (\ref{eq:aoki0}) in both orders.
The behavior in Eq.~(\ref{eq:aoki0}) may be arranged for by imposing the
generalized version \cite{Kniehl:2013uwa,Kniehl:2014} of the on-shell WFR
conditions \cite{Aoki:1982ed},
\begin{eqnarray}
\{[\hat{P}(\slashed{p})]^{-1}\}_{ij}u(\vec{p},M_j)&=&0,
\nonumber\\
\bar{u}(\vec{p},M_i)\{[\hat{P}(\slashed{p})]^{-1}\}_{ij}&=&0,
\nonumber\\
\left\{\frac{1}{\slashed{p}-M_i}\{[\hat{P}(\slashed{p})]^{-1}\}_{ii}\right\}
u(\vec{p},M_i)&=&u(\vec{p},M_i),
\nonumber\\
\bar{u}(\vec{p},M_i)
\left\{\{[\hat{P}(\slashed{p})]^{-1}\}_{ii}\frac{1}{\slashed{p}-M_i}\right\}
&=&\bar{u}(\vec{p},M_i),
\label{eq:aoki}
\end{eqnarray}
for $i,j=1,\ldots,N$.
Here, $u(\vec{p},M_i)$ and $\bar{u}(\vec{p},M_i)$ are four-component
spinors satisfying the Dirac equations,
\begin{equation}
(\slashed{p}-M_i)u(\vec{p},M_i)=\bar{u}(\vec{p},M_i)(\slashed{p}-M_i)=0.
\end{equation}
For stable Dirac fermions, an explicit proof that Eq.~(\ref{eq:aoki}) entails
Eq.~(\ref{eq:unit}) may be found in Sec.~III of Ref.~\cite{Kniehl:2012zb}.
This proof carries over to unstable Dirac fermions, as explained in Sec.~VII of
Ref.~\cite{Kniehl:2014}, and also to unstable Majorana fermions.

The WFR matrices $Z_\pm^{\nicefrac{1}{2}}$ may be determined by inserting
Eq.~(\ref{eq:reninv}) into Eq.~(\ref{eq:aoki}) and proceeding along the lines
of Refs.~\cite{Kniehl:2013uwa,Kniehl:2014}.
The results may be inferred from Refs.~\cite{Kniehl:2013uwa,Kniehl:2014} via
the substitution in Eq.~(\ref{eq:sub}).
Using also the last equality in Eq.~(\ref{eq:transpose}), we may translate
Eqs.~(57), (58), (71), and (72) of Ref.~\cite{Kniehl:2014} as
\begin{eqnarray}
(Z_\mp^{\nicefrac{1}{2}})_{ii}(Z_\pm^{\nicefrac{1}{2}})_{ii}&=&
\frac{M_{ii}^\mp M_{ii}^\pm}{s_i^\pm(M_i^2)[1-f_i^\prime(M_i^2)]},
\label{eq:zzs}\\
(Z_\pm^{\nicefrac{1}{2}})_{ii}^2&=&
\frac{M_i(M_{ii}^\pm)^2}{t_i^\pm(M_i^2)[1-f_i^\prime(M_i^2)]},
\label{eq:zzt}\\
(Z_\pm^{\nicefrac{1}{2}})_{ij}&=&\frac{M_{ij}^\pm}{M_{jj}^\pm}
(Z_\pm^{\nicefrac{1}{2}})_{jj},
\label{eq:z3}
\end{eqnarray}
where
\begin{eqnarray}
s_i^\pm(p^2)&=&[\bar{M}^\pm S_\pm(p^2)M^\pm]_{ii},
\label{eq:sl}\\
t_i^\pm(p^2)&=&[\bar{M}^\mp T_\pm(p^2)M^\pm]_{ii},
\label{eq:tl}\\
f_i(p^2)&=&\frac{t_i^+(p^2)t_i^-(p^2)}{s_i^+(p^2)s_i^-(p^2)},
\label{eq:fi}
\end{eqnarray}
and $M_{ij}^\pm$ and $\bar{M}_{ij}^\pm$ are defined in Eq.~(\ref{eq:m}).
Furthermore, Eq.~(65) of Ref.~\cite{Kniehl:2014} carries over as is,
\begin{equation}
M_i^2=f_i(M_i^2).
\label{eq:mass}
\end{equation}
As anticipated in the context of Eq.~(\ref{eq:mfe}), Eq.~(\ref{eq:mass})
provides an alternative all-order expression for $M_i$.
With the aid of the third and last equalities in Eq.~(\ref{eq:transpose}), we
observe that the two expressions in Eq.~(\ref{eq:sl}) actually coincide, so
that we may omit the superscript $\pm$,
\begin{equation}
s_i(p^2)=s_i^\pm(p^2),
\label{eq:sl1}
\end{equation}
and rewrite Eq.~(\ref{eq:fi}) as
\begin{equation}
f_i(p^2)=\frac{t_i^+(p^2)t_i^-(p^2)}{[s_i(p^2)]^2}.
\label{eq:fi1}
\end{equation}
Furthermore, we notice that Eqs.~(\ref{eq:zzs}) and (\ref{eq:zzt}) are
redundant.
In fact, Eq.~(\ref{eq:zzs}) follows from Eq.~(\ref{eq:zzt}) with the help of
Eqs.~(\ref{eq:fi}) and (\ref{eq:mass}).
We conclude that, owing to the Majorana-induced constraint in
Eq.~(\ref{eq:wfrct}), Eqs.~(\ref{eq:zzs})--(\ref{eq:z3}) uniquely determine the
WFR matrices to be
\begin{equation}
(Z_\pm^{\nicefrac{1}{2}})_{ij}=M_{ij}^\pm\left(
\frac{M_j}{t_j^\pm(M_j^2)[1-f_j^\prime(M_j^2)]}\right)^{\nicefrac{1}{2}}.
\label{eq:z}
\end{equation}
In contrast, the renormalization conditions in Eq.~(\ref{eq:aoki}) leave some
residual freedom in the determination of the WFR matrices for unstable Dirac
fermions, as explained in Refs.~\cite{Kniehl:2013uwa,Kniehl:2014}.

From Eqs.~(\ref{eq:mren}) and (\ref{eq:mass}), we obtain the all-order mass
counterterm as
\begin{equation}
\delta M_i=m_i^0-\sqrt{f_i(M_i^2)}.
\label{eq:mct}
\end{equation}
Alternatively, we could have used Eq.~(\ref{eq:mfe}) instead of
Eq.~(\ref{eq:mass}).
Using also Eq.~(\ref{eq:real}) and taking real and imaginary parts, we have
\begin{eqnarray}
m_i&=&\re\sqrt{f_i(M_i^2)}=m_i^0-\re\delta M_i,
\label{eq:realmass}\\
-\frac{\Gamma_i}{2}&=&\im\sqrt{f_i(M_i^2)}=-\im\delta M_i,
\label{eq:width}
\end{eqnarray}
where we have taken into account that the bare masses $m_i^0$ are real.
By the same token, the imaginary part of $\delta M_i$ is UV finite, as is
evident from Eq.~(\ref{eq:width}).

\section{WFR bifurcation}
\label{sec:five}

Let us assume temporarily that all the Majorana fermions are stable, with
$\Gamma_i=0$ in Eq.~(\ref{eq:real}).
In the complex $p^2$ plane, their mass shells $p^2=m_i^2$ are then all located
on the real axis below the thresholds of $[A_\pm(p^2)]_{ij}$ and
$[B_\pm(p^2)]_{ij}$, where the absorptive parts of the latter vanish.
Then, up to a sign flip in the $i\epsilon$ prescription, which is irrelevant
at this stage, the bare propagator matrix satisfies the pseudo-Hermiticity
condition $\gamma^0[P(\slashed{p})]^\dagger\gamma^0=P(\slashed{p})$
\cite{Aoki:1982ed},\footnote{%
On the right-hand side of this equation, we omitted the additional term
$i\int d^4x\,e^{ip\cdot x}$
$\times\langle0|[\Psi^0(x),[\Psi^0(0)]^TC^*]|0\rangle$.
In the noninteracting theory, its matrix elements in generation space,
$\delta_{ij}\left(\frac{1}{\slashed{p}-m_i^0-i\epsilon}-
\frac{1}{\slashed{p}-m_i^0+i\epsilon}\right)$, just flip the sign of the
$i\epsilon$ term in
$[P(\slashed{p})]_{ij}=\frac{\delta_{ij}}{\slashed{p}-m_i^0+i\epsilon}$.}
which implies via Eq.~(\ref{eq:bareinv}) that
$\gamma^0[\Sigma(\slashed{p})]^\dagger\gamma^0=\Sigma(\slashed{p})$
\cite{Espriu:2002xv}.
In turn, this implies via Eqs.~(\ref{eq:sigma}), (\ref{eq:stbare}),
(\ref{eq:cdbare}), (\ref{eq:m}), (\ref{eq:tl}), (\ref{eq:sl1}), and
(\ref{eq:fi1}) that 
\begin{eqnarray}
A_\pm^\dagger(p^2)&=&A_\mp(p^2),\qquad
B_\pm^\dagger(p^2)=B_\pm(p^2),\qquad
S_\pm^\dagger(p^2)=S_\pm(p^2),\qquad
T_\pm^\dagger(p^2)=T_\mp(p^2),
\nonumber\\
C_\pm^\dagger(p^2)&=&D_\pm(p^2),\qquad
E_\pm^\dagger(p^2)=F_\pm(p^2),\qquad
(M^\pm)^\dagger=\bar{M}^\pm,
\nonumber\\
{[s_i(p^2)]}^*&=&s_i(p^2),\qquad
[t_i^\pm(p^2)]^*=t_i^\mp(p^2),\qquad
[f_i(p^2)]^*=f_i(p^2),
\label{eq:reality}
\end{eqnarray}
where we have used $\adj(A^\dagger)=(\adj A)^\dagger$, which follows from
footnote~\ref{foot}, in the seventh equality.
Combining Eq.~(\ref{eq:reality}) with Eq.~(\ref{eq:transpose}), we find
\begin{eqnarray}
A_\pm^*(p^2)&=&A_\mp(p^2),\qquad
B_\pm^*(p^2)=B_\mp(p^2),\qquad
S_\pm^*(p^2)=S_\mp(p^2),\qquad
T_\pm^*(p^2)=T_\mp(p^2),
\nonumber\\
C_\pm^*(p^2)&=&C_\mp(p^2),\qquad
D_\pm^*(p^2)=D_\mp(p^2),\qquad
E_\pm^*(p^2)=E_\mp(p^2),\qquad
F_\pm^*(p^2)=F_\mp(p^2),
\nonumber\\
(M^\pm)^*&=&M^\mp,\qquad
(\bar{M}^\pm)^*=\bar{M}^\mp.
\label{eq:complex}
\end{eqnarray}
The first two equalities in Eq.~(\ref{eq:complex}) are in agreement with
Eq.~(4.6) of Ref.~\cite{Kniehl:1996bd}.
Using the last two equalities in Eq.~(\ref{eq:reality}) and the one before the
last in Eq.~(\ref{eq:complex}), we obtain from Eq.~(\ref{eq:z}) that
$Z_\pm^{\dagger\nicefrac{1}{2}}=Z_\mp^{T\nicefrac{1}{2}}$ \cite{Kniehl:1996bd},
which implies that Eqs.~(\ref{eq:wfrc}) and (\ref{eq:wfrct}) are related as
\begin{equation}
\bar{Z}^{\nicefrac{1}{2}}=\gamma^0Z^{\dagger\nicefrac{1}{2}}\gamma^0.
\label{eq:fake}
\end{equation}
From Eqs.~(\ref{eq:wfrct}) and (\ref{eq:fake}), it follows that
$\gamma^0CZ^{*\nicefrac{1}{2}}C^\dagger\gamma^0=Z^{\nicefrac{1}{2}}$, so that\break
$\Psi(x)=C\{[\Psi(x)]^\dagger\gamma^0\}^T$.

We now return to the general case of unstable Majorana fermions, with
$\Gamma_i>0$ in Eq.~(\ref{eq:real}).
In general, we then have
$\gamma^0[\Sigma(\slashed{p})]^\dagger\gamma^0\ne\Sigma(\slashed{p})$, so that
Eqs.~(\ref{eq:reality}) and (\ref{eq:complex}) no longer hold true, which
enforces the departure from Eq.~(\ref{eq:fake}).
Similar observations were made for unstable Dirac fermions at the one-loop
order in Ref.~\cite{Espriu:2002xv} and to all orders in
Refs.~\cite{Kniehl:2013uwa,Kniehl:2014}, where the notion WFR bifurcation was
coined.

\section{Two-loop results}
\label{sec:six}

To explore the anatomy of the all-order expressions for the
renormalization constants $\delta M_i$ and $(Z_\pm^{\nicefrac{1}{2}})_{ij}$
given in closed form in Eqs.~(\ref{eq:mct}) and (\ref{eq:z}), respectively, it
is useful to perform a perturbative expansion in the generic coupling constant
$\alpha$.
In Ref.~\cite{Kniehl:1996bd}, $\delta M_i$ and $(Z_\pm^{\nicefrac{1}{2}})_{ij}$
were expressed in terms of the self-energy functions $[A_\pm(p^2)]_{ij}$ and
$[B_\pm(p^2)]_{ij}$ at the one-loop order $\mathcal{O}(\alpha)$ in a general
renormalizable quantum field theory involving a mixed system of stable Majorana
fermions.
In the following, we assume $[A_\pm(p^2)]_{ij}$ and $[B_\pm(p^2)]_{ij}$ to be
known through the two-loop order $\mathcal{O}(\alpha^2)$ and allow for the
Majorana fermions to be unstable.
Our goal is to express $\delta M_i$ and $(Z_\pm^{\nicefrac{1}{2}})_{ij}$ in
terms of $[A_\pm(p^2)]_{ij}$ and $[B_\pm(p^2)]_{ij}$ through
$\mathcal{O}(\alpha^2)$.
In this way, we shall recover the well-known $\mathcal{O}(\alpha)$ results
\cite{Kniehl:1996bd} and present the $\mathcal{O}(\alpha^2)$ ones for the first
time.

Expanding Eqs.~(\ref{eq:fi1}) and (\ref{eq:zzt}) and the factor
$M_{ji}^\pm/M_{ii}^\pm$ for $j\ne i$ in Eq.~(\ref{eq:z3}) through
$\mathcal{O}(\alpha^2)$, we find
\begin{eqnarray}
f_i(p^2)&=&\frac{[T_+(p^2)]_{ii}[T_-(p^2)]_{ii}}{[S(p^2)]_{ii}^2}
+m_i^0[\tau_i^+(p^2)+\tau_i^-(p^2)-2m_i^0\sigma_i(p^2)]
+\mathcal{O}(\alpha^3),
\label{eq:fi2}\\
\frac{1}{(Z_\pm^{\nicefrac{1}{2}})_{ii}^2}&=&\frac{1-f_i^\prime(M_i^2)}{M_i}
\{[T_\pm(M_i^2)]_{ii}+\tau_i^\pm(M_i^2)\}
+\mathcal{O}(\alpha^3),
\label{eq:zzt2}\\
\frac{M_{ji}^\pm}{M_{ii}^\pm}&=&\frac{1}{M_i^2-M_j^2}
\left\{[F_\pm(M_i^2)]_{ji}
\left\{1+\frac{[F_\pm(M_i^2)]_{jj}-M_j^2}{M_i^2-M_j^2}\right\}
\vphantom{\sum_{i\ne k\ne j}}\right.
\nonumber\\
&&{}+\left.\sum_{i\ne k\ne j}\frac{[F_\pm(M_i^2)]_{jk}[F_\pm(M_i^2)]_{ki}}
{M_i^2-M_k^2}\right\}
+\mathcal{O}(\alpha^3)
\qquad(j\ne i),
\label{eq:z2}
\end{eqnarray}
respectively, where we have exploited the third equality in
Eq.~(\ref{eq:transpose}) to introduce $[S(p^2)]_{ii}=[S_\pm(p^2)]_{ii}$ and
\begin{eqnarray}
\tau_i^\pm(p^2)&=&\sum_{j\ne i}\frac{[F_\pm(M_i^2)]_{ji}}{M_i^2-M_j^2}
\left\{m_j^0\frac{[F_\pm(M_i^2)]_{ji}}{M_i^2-M_j^2}+2[A_\pm(p^2)]_{ji}\right\},
\nonumber\\
\sigma_i(p^2)&=&\sum_{j\ne i}\frac{1}{M_i^2-M_j^2}
\left\{\frac{[F_+(M_i^2)]_{ji}[F_-(M_i^2)]_{ji}}{M_i^2-M_j^2}
-[F_+(M_i^2)]_{ji}[B_-(p^2)]_{ji}\right.
\nonumber\\
&&{}-\left.\vphantom{\frac{[F_+(M_i^2)]_{ji}[F_-(M_i^2)]_{ji}}{M_i^2-M_j^2}}
[F_-(M_i^2)]_{ji}[B_+(p^2)]_{ji}\right\}.
\label{eq:ts}
\end{eqnarray}
At this point, a few comments are in order.
For $j\ne i$, $(Z_\pm^{\nicefrac{1}{2}})_{ji}$ may be evaluated by substituting
Eqs.~(\ref{eq:zzt2}) and (\ref{eq:z2}) into Eq.~(\ref{eq:z3}).
The first term on the right-hand side of Eq.~(\ref{eq:fi2}) is the contribution
that survives if the intergeneration mixing is turned off.
For the sake of a compact notation, it is written in a factorized form, which
is to be expanded through $\mathcal{O}(\alpha^2)$ to become
\begin{eqnarray}
\frac{[T_+(p^2)]_{ii}[T_-(p^2)]_{ii}}{[S(p^2)]_{ii}^2}
&=&m_i^0\{m_i^0+[A_+(p^2)]_{ii}+[A_-(p^2)]_{ii}\}\{1+2[B(p^2)]_{ii}\}
\nonumber\\
&&{}+[A_+(p^2)]_{ii}[A_-(p^2)]_{ii}+3(m_i^0)^2[B(p^2)]_{ii}^2
+\mathcal{O}(\alpha^3),
\label{eq:no}
\end{eqnarray}
where we have used the second equality in Eq.~(\ref{eq:transpose}) to define
$[B(p^2)]_{ii}=[B_\pm(p^2)]_{ii}$.
In Eq.~(\ref{eq:no}), it is understood that, in products of two loop functions,
each factor is to be evaluated at $\mathcal{O}(\alpha)$, while loop functions
that do not appear in such products are to be evaluated through
$\mathcal{O}(\alpha^2)$.
This also applies to Eqs.~(\ref{eq:fi2})--(\ref{eq:ts}).

To express the mass counterterms $\delta M_i$ in Eq.~(\ref{eq:mct})
and the WFR matrix elements $(Z_\pm^{\nicefrac{1}{2}})_{ji}$ in
Eq.~(\ref{eq:z}) in terms of renormalized parameters, we may proceed as
follows.
We first evaluate the right-hand side of Eq.~(\ref{eq:mct}) through
$\mathcal{O}(\alpha^2)$ using Eq.~(\ref{eq:fi2}) in combination with
Eqs.~(\ref{eq:ts}) and (\ref{eq:no}).
Since $\delta M_i$ starts at $\mathcal{O}(\alpha)$, it is sufficient to
eliminate $m_i^0$ on the right-hand side of Eq.~(\ref{eq:mct}) using
Eq.~(\ref{eq:mren}) with $\delta M_i$ evaluated to $\mathcal{O}(\alpha)$.
The latter may be read off from Eqs.~(\ref{eq:fi2}) and (\ref{eq:no}) and reads
\begin{equation}
\delta M_i=-\frac{I_i(M_i^2)}{2M_i}+\mathcal{O}(\alpha^2),
\label{eq:mct1}
\end{equation}
with
\begin{equation}
I_i(p^2)=M_i\{[A_+(p^2)]_{ii}+[A_-(p^2)]_{ii}\}+2M_i^2[B(p^2)]_{ii},
\label{eq:ii}
\end{equation}
where we have replaced $m_i^0$ by $M_i$, with no effect to the order
considered.
By the same token, appearances of $m_i^0$ in the $\mathcal{O}(\alpha^2)$ term
on the right-hand side of Eq.~(\ref{eq:mct}) may be replaced by $M_i$.
Explicit and implicit appearances of $m_j^0$ with $j\ne i$ at
$\mathcal{O}(\alpha)$ and $\mathcal{O}(\alpha^2)$ may be eliminated in the same
way as those of $m_i^0$.
The implicit dependence on $m_j^0$ of an $\mathcal{O}(\alpha)$ quantity,
$f(m_j^0)$ say, is conveniently eliminated by Taylor expansion as
$f(m_j^0)=f(M_j)+\delta M_j\partial f(M_j)/\partial M_j+\mathcal{O}(\alpha^3)$,
where $\delta M_j$ is calculated to $\mathcal{O}(\alpha)$ from
Eq.~(\ref{eq:mct1}).
In general, the resulting $\mathcal{O}(\alpha^2)$ expression for $\delta M_i$
still implicitly depends on other bare parameters, such as boson masses,
coupling constants, and mixing-matrix elements, which also require
renormalization.
After this, $m_i^0$ is expressed via Eq.~(\ref{eq:mren}) through
$\mathcal{O}(\alpha^2)$ entirely in terms of renormalized parameters and may
thus be eliminated from Eqs.~(\ref{eq:zzt2}) and (\ref{eq:z2}).
Appearances of $m_j^0$ with $j\ne i$ and other bare parameters are eliminated
from these equations as explained above for Eq.~(\ref{eq:mct}).

For $p^2=M_i^2$, there are some cancellations in the function $f_i(p^2)$ given
by Eqs.~(\ref{eq:fi2}), (\ref{eq:ts}), and (\ref{eq:no}) through
$\mathcal{O}(\alpha^2)$, yielding
\begin{eqnarray}
f_i(M_i^2)&=&\frac{[T_+(M_i^2)]_{ii}[T_-(M_i^2)]_{ii}}{[S(M_i^2)]_{ii}^2}
+M_i\sum_{j\ne i}\left\{\frac{[F_+(M_i^2)]_{ji}}{M_i^2-M_j^2}
\left\{[A_+(M_i^2)]_{ji}+M_i[B_-(M_i^2)]_{ji}\right\}
\right.
\nonumber\\
&&{}+\left.
\vphantom{\frac{[F_+(M_i^2)]_{ji}}{M_i^2-M_j^2}}
(+\leftrightarrow-)\right\}
+\mathcal{O}(\alpha^3).
\label{eq:fim}
\end{eqnarray}
The quantity $f_i^\prime(M_i^2)$ appearing in Eq.~(\ref{eq:zzt2}) is required
through $\mathcal{O}(\alpha^2)$.
Through this order, it may be conveniently evaluated as
\begin{eqnarray}
f_i^\prime(M_i^2)&=&M_i^2\left\{
\frac{[A_+^\prime(M_i^2)]_{ii}}{[T_+(M_i^2)]_{ii}}
+\frac{[A_-^\prime(M_i^2)]_{ii}}{[T_-(M_i^2)]_{ii}}
+2\frac{[B^\prime(M_i^2)]_{ii}}{[S(M_i^2)]_{ii}}\right\}
+2M_i\sum_{j\ne i}\left\{\frac{[F_+(M_i^2)]_{ji}}{M_i^2-M_j^2}\right.
\nonumber\\
&&{}\times\left.
\vphantom{\frac{[F_+(M_i^2)]_{ji}}{M_i^2-M_j^2}}
\left\{[A_+^\prime(M_i^2)]_{ji}+M_i[B_-^\prime(M_i^2)]_{ji}\right\}
+(+\leftrightarrow-)\right\}
+\mathcal{O}(\alpha^3),
\label{eq:fip}
\end{eqnarray}
where we have used Eq.~(\ref{eq:mass}) through $\mathcal{O}(\alpha)$ to
eliminate the combination on the left-hand side of Eq.~(\ref{eq:no}).
The functions $[F_\pm(p^2)]_{ji}$ with $j\ne i$ are required through
$\mathcal{O}(\alpha^2)$ at their first appearance in Eq.~(\ref{eq:z2}) and
through $\mathcal{O}(\alpha)$ elsewhere in Eqs.~(\ref{eq:z2}), (\ref{eq:ts}),
(\ref{eq:fim}), and (\ref{eq:fip}).
Through $\mathcal{O}(\alpha^2)$, we have
\begin{eqnarray}
[F_\pm(p^2)]_{ji}&=&M_j\{[I_N+B_\mp(p^2)]A_\pm(p^2)\}_{ji}
+[A_\mp(p^2)+A_\mp(p^2)B_\mp(p^2)+B_\pm(p^2)A_\mp(p^2)]_{ji}M_i
\nonumber\\
&&{}+\{B_\pm(p^2)[I_N+B_\pm(p^2)]\}_{ji}M_i^2
+M_j\{B_\mp(p^2)[I_N+B_\mp(p^2)]\}_{ji}M_i
\nonumber\\
&&{}+[A_\mp(p^2)A_\pm(p^2)]_{ji}
+\sum_k[B_\pm(p^2)]_{jk}M_k\{[A_\pm(p^2)]_{ki}+[B_\mp(p^2)]_{ki}M_i\}
\nonumber\\
&&{}+\delta M_j\{[A_\pm(p^2)]_{ji}+[B_\mp(p^2)]_{ji}M_i\}
+\{[A_\mp(p^2)]_{ji}+M_j[B_\mp(p^2)]_{ji}
\nonumber\\
&&{}+2[B_\pm(p^2)]_{ji}M_i\}\delta M_i
+\mathcal{O}(\alpha^3).
\end{eqnarray}
Finally, the combination $[F_\pm(M_i^2)]_{jj}-M_j^2$ appearing in
Eq.~(\ref{eq:z2}) is required through $\mathcal{O}(\alpha)$, where it may be
rewritten in terms of the function $I_i(p^2)$ in Eq.~(\ref{eq:ii}) as
\begin{equation}
[F_\pm(M_i^2)]_{jj}-M_j^2=I_j(M_i^2)-I_j(M_j^2)
+\mathcal{O}(\alpha^2).
\label{eq:com}
\end{equation}

In the case of stable Majorana fermions, in which Eqs.~(\ref{eq:reality}) and
(\ref{eq:complex}) apply, the $\mathcal{O}(\alpha)$ terms of
Eqs.~(\ref{eq:zzt}), (\ref{eq:z3}), and (\ref{eq:mct}) evaluated using
Eqs.~(\ref{eq:fi2})--(\ref{eq:z2}) and (\ref{eq:no}) agree with Eqs.~(4.11),
(4.10), and (4.12) of Ref.~\cite{Kniehl:1996bd}, respectively.

\section{Conclusions}
\label{sec:seven}

We renormalized the propagator matrix of a mixed system of unstable Majorana
fermions in a general parity-nonconserving quantum field theory adopting the
pole scheme, in which the pole masses serve as the renormalized masses.
The squares of the pole masses are the complex poles of the propagator matrix.
The inverse propagator matrix is built up by the one-particle-irreducible
Feynman diagrams pertaining to the transitions of fermion $j$ to fermion $i$
order by order in perturbation theory.
In gauge theories, the pole masses are expected to be gauge independent.
This was proven for the SM \cite{Gambino:1999ai} using Nielsen identities
\cite{Nielsen:1975fs}.
In spontaneously broken gauge theories, one needs to include the tadpoles to
ensure the gauge independence of the mass counterterms.
This then carries over to the pole masses because the bare masses are gauge
independent as a matter of principle.

The WFR matrices were determined by requiring that each diagonal element of the
renormalized propagator matrix resonates with unit residue if the respective
fermion is on its mass shell.
%This renormalization condition avoids finite renormalizations that are
%otherwise required by the LSZ reduction formalism \cite{Lehmann:1954rq}.
This renormalization condition is singled out by the LSZ reduction formalism
\cite{Lehmann:1954rq} because it avoids finite renormalizations that are
otherwise required.
In this sense, it may be considered scheme independent.
Furthermore, it uniquely fixes the WFR matrices.
This is in contrast to the Dirac case, where some residual freedom exists,
which may be exhausted by imposing an additional WFR condition
\cite{Kniehl:2013uwa,Kniehl:2014,Kniehl:1996bd}.
Specifically, this residual freedom affects the pairs of WFR matrix elements
that appear as factors in the off-diagonal entries of the renormalized
propagator matrix.
As for Dirac fermions \cite{Kniehl:2013uwa,Kniehl:2014,Espriu:2002xv}, we
encountered WFR bifurcation in the case of instability, i.e., the WFR matrices
of the in and out states are no longer related by Hermitian conjugation, so
that Eq.~(\ref{eq:fake}) is violated.
However, they are still related by Eq.~(\ref{eq:wfrct}), which is a consequence
of the Majorana condition in Eq.~(\ref{eq:majom}) and is absent in the Dirac
case.

The dressed propagator matrix and the renormalization constants are expressed
in terms of the unrenormalized self-energies of the $j\to i$ transitions, which
have scalar, pseudoscalar, vector, and axial vector parts.
Owing to the Majorana condition in Eq.~(\ref{eq:majom}), the latter are subject
to the symmetry relations given by the first two equalities in
Eq.~(\ref{eq:transpose}).
We presented closed analytic results, which are valid to all orders because we
refrained from explicit perturbative expansions.
Specifically, the renormalized dressed propagator matrix is given by
Eq.~(\ref{eq:renz}), the pole mass counterterms by Eq.~(\ref{eq:mct}), and the
WFR matrices by Eq.~(\ref{eq:z}).
In these formulas, the renormalized masses $M_i$ enter as arguments $p^2=M_i^2$
of the various self-energy functions, and it is understood that the latter are
evaluated from the bare Lagrangian of the considered quantum field theory, so
that the masses, couplings, and mixing angles on which they depend are all bare
parameters to start with.

Apart from being interesting in their own right, the results presented here
have a number of important phenomenological applications.
In the following, we mention but three of them.
First, in the perturbative treatment of a specific particle scattering or decay
process involving stable or unstable Majorana fermions, our formulas for
$\delta M_i$ and $(Z_\pm^{\nicefrac{1}{2}})_{ij}$ may be used after expansion
through the considered order and truncation of terms beyond that order.
If the unstable Majorana fermion $i$ occurs on an internal line, then
$\delta M_i$ enters.
If it occurs on an external line, then $(Z_\pm^{\nicefrac{1}{2}})_{ij}$ enters.
Strictly speaking, unstable particles are not entitled to appear in asymptotic
states of scattering amplitudes in quantum field theory.
However, in numerous applications of significant phenomenological interest, the
rigorous compliance with this tenet would immediately entail a proliferation of
external legs and bring the evaluation of radiative corrections to a grinding
halt, the more so as almost all the known elementary particles are unstable.
For the reader's convenience, we presented explicit two-loop expressions
for the renormalization constants, in Eqs.~(\ref{eq:fi2})--(\ref{eq:com}),
which may be employed in phenomenological applications involving Majorana
fermions as they stand.
In the one-loop case of stable Majorana fermions, Eqs.~(4.10)--(4.12) in
Ref.~\cite{Kniehl:1996bd} are reproduced by Eqs.~(\ref{eq:z3}), (\ref{eq:zzt}),
and (\ref{eq:mct}), respectively, evaluated at $\mathcal{O}(\alpha)$ using
Eqs.~(\ref{eq:fi2})--(\ref{eq:z2}) and (\ref{eq:no}).

Second, the total decay widths $\Gamma_i$ may be perturbatively evaluated
through any order from the unrenormalized self-energy functions
$[A_\pm(p^2)]_{ij}$ and $[B_\pm(p^2)]_{ij}$ by solving Eq.~(\ref{eq:width})
iteratively.

Third, our result for the mass counterterms $\delta M_i$ may be used to
switch from the pole scheme adopted here to any other scheme of mass
renormalization, as long as the method of regularization is maintained. 
In fact, since the bare masses $m_i^0$ are independent of the choice of
renormalization scheme, the equivalent of Eq.~(\ref{eq:mren}) in some other
scheme is
\begin{equation}
m_i^0=\tilde{M}_i+\delta\tilde{M}_i,
\label{eq:mren3}
\end{equation}
where $\delta\tilde{M}_i$ has the same UV singularities as $\delta M_i$, but
differs in the finite terms.
In gauge theories, preferable renormalization schemes are those in which
$\delta\tilde{M}_i$ is arranged to be gauge independent, so that $\tilde{M}_i$
enjoys the same desirably property.
Equating Eqs.~(\ref{eq:mren}) and (\ref{eq:mren3}), we thus obtain a UV-finite
relationship between the renormalized masses of both schemes,
\begin{equation}
M_i=\tilde{M}_i+\delta\tilde{M}_i-\delta M_i.
\label{eq:change}
\end{equation}
A quantity evaluated to a given order of perturbation theory in the pole scheme
may then be translated to the other mass renormalization scheme by substituting
Eq.~(\ref{eq:change}), expanding in the coupling constant, and discarding terms
beyond the considered order.

In the context of perturbative calculations in quantum chromodynamics, the
modified minimal-subtraction ($\overline{\mathrm{MS}}$) scheme
\cite{Bardeen:1978yd} of dimensional regularization \cite{Bollini:1972ui} is
frequently employed in the literature.
A natural extension of the $\overline{\mathrm{MS}}$ definition of mass to the
EW sector of the SM may be obtained from Eq.~(\ref{eq:realmass}) by writing
\begin{eqnarray}
m_i^0&=&m_i+\re\delta M_i
\nonumber\\
&=&\overline{m}_i+\delta\overline{m}_i,
\label{eq:msbar}
\end{eqnarray}
where
\begin{equation}
\delta\overline{m}_i=(\re\delta M_i)_\mathrm{UV}
\label{eq:msbar1}
\end{equation}
collects just the poles in $\varepsilon=2-d/2$, with $d$ being the
dimensionality of space-time, and the familiar terms involving
$\gamma_E-\ln(4\pi)$, with $\gamma_E$ being Euler's constant, that appear
in $\re\delta M_i$ order by order.
The latter may be absorbed by an appropriate redefinition of the
renormalization scale $\mu$, namely,
$\mu=\mu^\prime\exp(\gamma_E/2)/(2\sqrt{\pi})$ \cite{Kniehl:1994ph}.
From Eqs.~(\ref{eq:msbar}) and (\ref{eq:msbar1}), it hence follows that
\begin{eqnarray}
m_i&=&\overline{m}_i+(\re\delta M_i)_\mathrm{UV}-\re\delta M_i
\nonumber\\
&=&\overline{m}_i-(\re\delta M_i)_{\overline{\mathrm{MS}}},
\label{eq:msbar2}
\end{eqnarray}
where $(\re\delta M_i)_{\overline{\mathrm{MS}}}$ is the UV-finite remainder of
$\re\delta M_i$ after $\overline{\mathrm{MS}}$ subtraction of the poles in
$\varepsilon$ at renormalization scale $\mu$. 
As mentioned above, it is necessary to include the tadpole contributions in
$(\re\delta M_i)_\mathrm{\overline{\mathrm{MS}}}$ in order for $\overline{m}_i$
to be gauge independent \cite{Hempfling:1994ar}.
Otherwise, the functional dependencies of radiatively corrected transition
matrix elements on such renormalized masses acquire artificial gauge
dependence, and the choices of gauge must always be specified along with the
values of such renormalized masses extracted from experimental data.
The necessity to include the tadpole contributions in order to render the
mass counterterms gauge independent was also noticed within the on-shell
renormalization of the SM at one loop
\cite{Fleischer:1980ub,Bohm:1986rj,Sirlin:1985ux}.
In that case, however, the omission of the tadpole contributions would be
inconsequential in practice, since the functional dependencies of radiatively
corrected transition matrix elements on the renormalized masses could be
preserved at the expense of allowing for the bare masses to become gauge
dependent.
Unfortunately, such an escape is unavailable in the case of
Eq.~(\ref{eq:msbar2}), which directly relates the mass definitions in two
different renormalization schemes \cite{Hempfling:1994ar}.
 
\section*{Acknowledegments}

We thank Alberto Sirlin for numerous valuable discussions and Concha
Gonzalez-Garcia for an illuminating communication regarding
Ref.~\cite{Gonzalez-Garcia:2013jma}.
This research was supported in part by the German Research Foundation through
the  Collaborative Research Center No.~SFB~676 {\it Particles, Strings and the
Early Universe---The Structure of Matter and Space Time}.

\end{document}